\documentstyle[emulateapj]{article}

\received{} 
\revised{} 
\accepted{} 
\cpright{}{} 
\journalid{}{} 
\articleid{}{} 
\paperid{} 
 
\cpright{}{} 
\ccc{} 
 
\slugcomment{Accepted by ApJ Letters}
 
\lefthead{} 
\righthead{} 
 
\begin{document} 

\title{Precession of collimated outflows from young stellar objects}
 
\author{C. Terquem\altaffilmark{1,4,5},
J. Eisl\"offel\altaffilmark{2}, J. C. B. Papaloizou\altaffilmark{3,4}
and R. P. Nelson\altaffilmark{3,4}}

\altaffiltext{1}{UCO/Lick Observatory, University of California,
Santa~Cruz, CA~95064~-- ct@ucolick.org}

\altaffiltext{2}{Th\"uringer Landessternwarte,
Karl--Schwarzschild--Observatorium, Sternwarte 5, D-07778~Tautenburg,
Germany~-- jochen@tls-tautenburg.de}

\altaffiltext{3}{Astronomy Unit, School of Mathematical Sciences,
Queen Mary~\& Westfield College, Mile End Road, London~E1~4NS, UK~--
J.C.B.Papaloizou@qmw.ac.uk, R.P.Nelson@qmw.ac.uk}

\altaffiltext{4}{Isaac Newton Institute for Mathematical Sciences,
University of Cambridge, 20 Clarkson Road, Cambridge~CB3~0EH, UK}

\altaffiltext{5}{On leave from: Laboratoire d'Astrophysique,
Observatoire de Grenoble, Universit\'e Joseph Fourier/CNRS,
38041~Grenoble Cedex~9, France}

\begin{abstract} 

We consider several protostellar systems where either a precessing jet
or at least two misaligned jets have been observed.  We assume that
the precession of jets is caused by tidal interactions in noncoplanar
binary systems. For Cep~E, V1331~Cyg and RNO~15--FIR the inferred
orbital separations and disk radii are in the range 4--160~AU and
1--80~AU, respectively, consistent with those expected for pre--main
sequence stars.  Furthermore, we assume or use the fact that the
source of misaligned outflows is a binary, and evaluate the
lengthscale over which the jets should precess as a result of tidal
interactions.  For T~Tau, HH1~VLA~1/2 and HH~24~SVS63, it may be
possible to detect a bending of the jets rather than 'wiggling'. In
HH~111~IRS and L1551~IRS5, 'wiggling' may be detected on the current
observed scale.  Our results are consistent with the existence of
noncoplanar binary systems in which tidal interactions induce jets to
precess.

\end{abstract} 
 
 
\section{Introduction} 

Most T~Tauri stars are observed to be in multiple systems
(Mathieu~1994).  There is also indirect evidence that the sizes of
circumstellar disks contained within binary systems are correlated
with the binary separation (Osterloh~\& Beckwith~1995, Jensen,
Mathieu~\& Fuller~1996).  This suggests that binary companions are
responsible for limiting the sizes of the discs through tidal
truncation (Papaloizou~\& Pringle~1977, Paczy\'nski~1977).  There are
indications that, in binaries, the plane of at least one of the
circumstellar discs and that of the orbit may not necessarily be
aligned.  The most striking evidence for such noncoplanarity is given
by HST and adaptive optics images of HK~Tau (Stapelfeldt et al. 1998,
Koresko~1998).  Also, observations of molecular outflows in star
forming regions commonly show several jets of different orientations
emanating from an unresolved region the extent of which can be as
small as $\sim 100$~AU (Davis, Mundt~\& Eisl\"offel~1994).  These jets
are usually believed to originate from a binary in which circumstellar
disks are misaligned, and in some of these systems a binary has indeed
been resolved.  In such cases, we expect tidal interaction to induce
the precession of the disk (possibly both disks) which is not in the
orbital plane, and thus of any jet it drives.  Furthermore, a number
of jets seem to be precessing (see below).  The above discussion leads
us to interpret this precessional motion as being driven by tidal
interactions between the disk from which the jet originates and a
companion on an inclined orbit.

In this {\it Letter} we consider several observed systems in the light
of this model.  In \S~\ref{sec:theory} we review the theory of
precessing warped disks. In \S~\ref{sec:obs} we apply it to observed
systems. We first consider cases where a precessing jet has been
observed and calculate the parameters of the binaries in which tidal
interactions would produce the observationally inferred precession
frequencies.  We next study cases where misaligned jets have been
observed and, assuming or using the fact that the source of these
outflows is a binary, we calculate the precession frequency that would
be induced by tidal interactions in the binary and the lengthscale
over which the jets should 'wiggle' or bend as a result of this
precessional motion.  In \S~\ref{sec:disc} we give a summary and
discussion of our results.

\section{Theory of precessing warped disks}
\label{sec:theory}

We consider a binary system in which the primary and the secondary
have a mass $M_p$ and $M_s$, respectively.  We suppose that the primary
is surrounded by a disk of radius $R$ with negligible mass so that
precession of the orbital plane can be neglected.  The binary orbit is
assumed circular with radius $D$ and is in a plane with inclination
angle $\delta$ to the plane of this disk.  In general, $\delta$ will
be evolving with time.  However, this evolution, {\it which is not
necessarily toward coplanarity}, occurs on a long timescale (equal or
larger than the disk viscous timescale) if the warp is not too severe
(Papaloizou~\& Terquem~1995), so that we will consider here $\delta$
as a constant.  This means that although the outer parts of the
disk may be driven out of the initial disk plane on a relatively short
timescale, the inner parts will retain their orientation with respect
to the orbital plane for a timescale comparable to the disk viscous
timescale.  Since jets are expected to originate from the disk inner
parts, their orientation relative to the orbital plane will be
determined by that of the disk inner regions. 

The secular perturbation caused by the companion leads to the
precession of the disk about the orbital axis, as in a gyroscope.
{\it The disk is expected to precess as a rigid body} if it can
communicate with itself through some physical process on a timescale
less than the precession period.  In non--self gravitating
protostellar disks, communication is governed by bending waves
(Papaloizou~\& Lin 1995, Larwood et al. 1996, Terquem 1998).  The
condition for rigid body precession is then satisfied if $H/r > \left|
\omega_p \right| / \Omega_0,$ where $\omega_p$ is the (uniform)
precession frequency in the disk and $\Omega_0$ is the angular
velocity at the disk outer edge (Papaloizou~\& Terquem 1995).  An
expression for $\omega_p$ has been derived by Papaloizou~\& Terquem
(1995). Here we just give an approximate expression, that we derive by
assuming that the disk surface density is uniform and that the
rotation is Keplerian:

\begin{equation}
\omega_p = - \frac{15}{32} \frac{M_s}{M_p} \left( \frac{R}{D}
\right)^3 \cos \delta \sqrt{\frac{G M_p}{R^3}},
\label{prec}
\end{equation}

\noindent where $G$ is the gravitational constant. The assumptions
under which this expression is valid (see Terquem 1998 for a summary)
are usually satisfied in the case of the relatively wide binaries we
will be discussing here and for the values of $D/R$ we will be
considering.  Although we have assumed a circular orbit, an eccentric
binary orbit can be considered by replacing $D$ by the semi-major axis
and multiplying the precession frequency by $(1-e^2)^{-3/2}$, where
$e$ is the eccentricity.  On this basis we consider that the
discussion presented below should remain valid for $0<e<1/2.$

\section{Application to particular systems}
\label{sec:obs}

\subsection{Expected parameters of binary systems with precessing jets}
\label{sec:obs1}

Observations of molecular outflows in star forming regions show in
some cases ``wiggling'' knots (or a helical pattern in projection onto
the plane of the sky), which can be interpreted as being the result of
the precession of the outflowing jet.  In this section we will assume
that such precession is caused by tidal interaction between the disk
from which the outflow originates and a companion star in a
noncoplanar orbit.  In cases where the outflow has lasted many
precession periods the implication is that the ``wiggling'' should be
periodic.  If we indeed assume that this is the case, the observations
give the projected wavelength $\lambda_{proj}.$ When the angle $i$
between the outflow and the line of sight can be estimated, the actual
wavelength $\lambda = \lambda_{proj} / \sin i$ can be derived.  The
precession period is then given by $T= \lambda / v$, where $v$ is the
outflow velocity, and $\left| \omega_p \right|= 2 \pi / T.$
Furthermore, if the outflow is precessing because the disk plane and
that of the orbit are misaligned, the angle $\delta$ between these two
planes is equal to the angle between the central flow axis and the
line of maximum deviation of the flow from this axis.  This angle can
also be observed. In all the cases studied in this section, $\delta$
is small enough ($\sim 10-20\arcdeg$) so that we will consider $\cos
\delta =1$.

In this section we will adopt $M_p = 0.5$~$M_{\sun}$.  Since we do not
know whether the jet which is observed to precess originates from the
primary or the secondary, we will consider mass ratios $M_s/M_p$
between 0.5 and 2. The lowest (largest) values would correspond to the
case where the jet originates from the primary (secondary).  These
values are typical for pre--main sequence binaries.  Finally, the disk
from which the outflow originates is expected to have its size
truncated by tidal interaction with the companion star in such a way
that $D/R$ lies between 2 and 4. Since Larwood et al.~(1996) have
shown that tidal truncation is only marginally affected by lack of
coplanarity, in this section we will consider $2 \le D/R \le
4$. Assuming a fixed ratio $R/D,$ equation~(\ref{prec}) can then be
used to calculate the disk radius:

\begin{equation}
R = \left( \frac{15}{32} \frac{M_s}{M_p} \cos \delta \frac{\sqrt{ G
M_p}}{\left| \omega_p \right|} \right)^{2/3} \left( \frac{R}{D}
\right)^2 .
\label{radius}
\end{equation}

\noindent Since $M_p$ appears with the power $1/3$, an uncertainty of
a factor 2 over $M_p$ is equivalent to an uncertainty of only a factor
1.26 over $R$. The main uncertainty over $R$ comes through the ratio
$R/D$. We now consider some particular protostellar systems in the
light of the above discussion.

\begin{figure}
\plotone{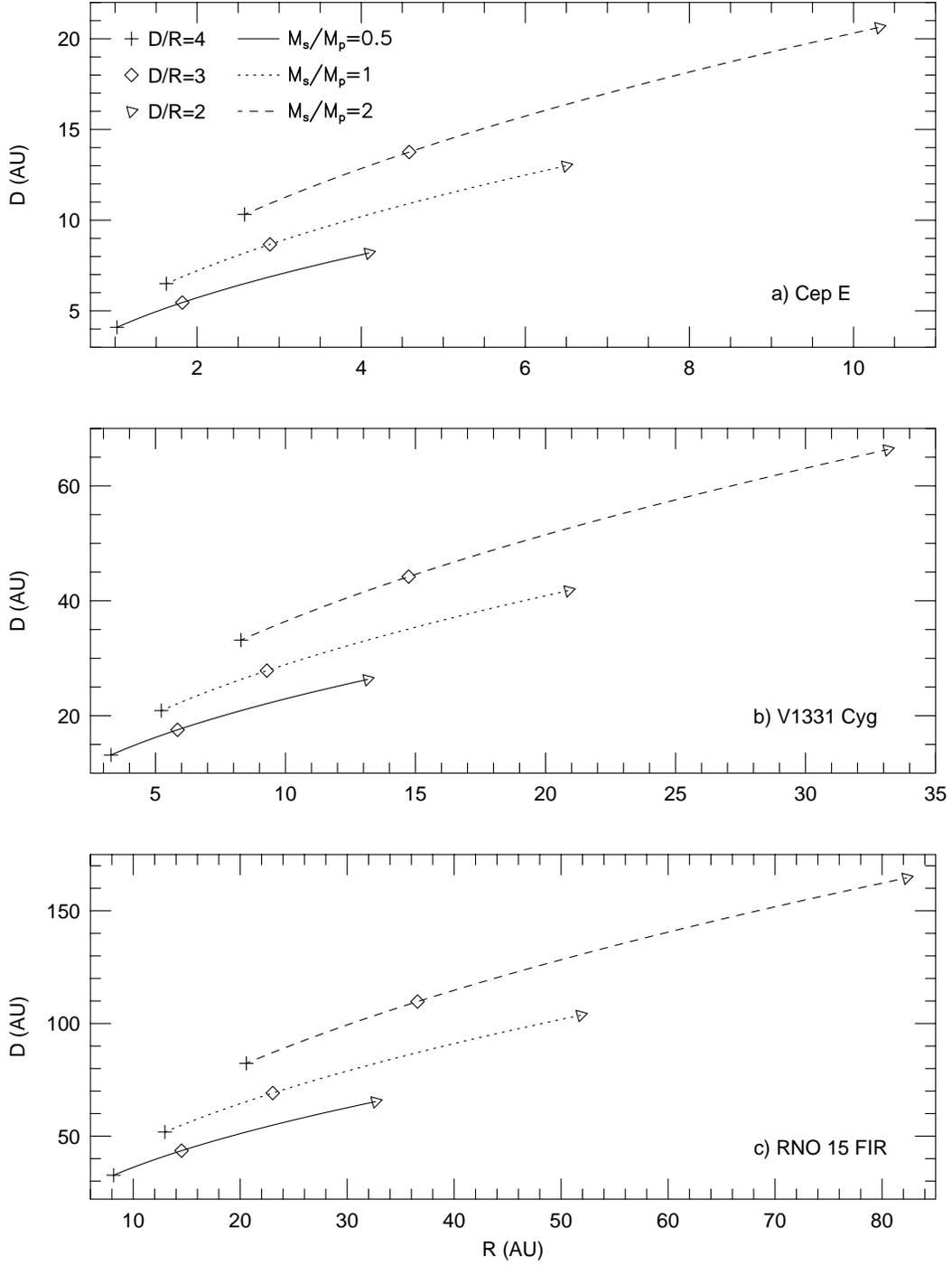}
\caption[]{ The binary separation $D$ vs. the disk radius $R$ in~AU, as
calculated from equation~(\ref{radius}), for Cep~E ({\it a}),
V1331~Gyg ({\it b}) and RNO~15--FIR ({\it c}). The different curves
correspond to $M_s/M_p=0.5$ ({\it solid line}), 1 ({\it dotted line})
and 2 ({\it dashed line}).  $D/R$ is kept in the range 2--4 and the
symbols indicate $D/R=4$ ({\it crosses}), 3 ({\it squares}) and 2
({\it triangles}). }
\label{fig1} 
\end{figure}

\noindent {\bf{Cep~E}}, at a distance of 730~pc, drives two outflows
almost perpendicular to each other Eisl\"offel~et al.~(1996), which
suggests that this source is a binary. In addition, they have
interpreted the morphology of one of these jets as due to precession,
and they have inferred $T=400$~years.  Figure~\ref{fig1}.a shows a
plot of $D$ against $R$, as calculated from equation~(\ref{radius}).
We see that $R$ lies in the range $1-10$~AU while $D$ lies in the
range $4-20$~AU.  Binary separation would be 0.''005 to 0.''03, which
is not currently resolvable.

\noindent {\bf{V1331~Cyg}} is located at a distance of 550~pc.
Visible line emission shows a very faint and diffuse feature in the
vicinity of this object, which appears to be a strongly 'wiggling' jet
(Mundt~\& Eisl\"offel~1998).  The observations give $\lambda_{proj}
\simeq 0.5$~pc (a full period is observed).  Using $i \sim 42 \arcdeg$
(Mundt~\& Eisl\"offel~1998), we derive $\lambda=0.71$~pc.  Since $v
\sim 300$~km\,s$^{-1}$ (Mundt~\& Eisl\"offel~1998), we get
$T=2,300$~years.  Figure~\ref{fig1}.b shows a plot of $D$ against $R$,
as calculated from equation~(\ref{radius}).  We see that $R$ lies in
the range $3-33$~AU while $D$ lies in the range $13-66$~AU.  Binary
separation would be 0.''02 to 0.''1. This upper value may be possible
to resolve with the VLA or adaptive optics in the near--infrared.

\noindent {\bf{RNO~15--FIR}}, located at a distance of 350~pc, drives
a molecular outflow which appears to be 'wiggling' (Davis~et
al.~1997).  It is possible to interpret this morphology as due to
precession within the uncertainty of the measurement (see Fig.~7 of
Davis~et al.~1997).  From the observations, we derive $\lambda_{proj}=
0.065$~pc.  Assuming $i=45 \arcdeg$ (Cabrit~1989) and
$v=10$~km\,s$^{-1}$ (Davis~et al.~1997), we get $\lambda=0.092$~pc and
$T=9,000$~years.  Figure~\ref{fig1}.c shows a plot of $D$ against $R$,
as calculated from equation~(\ref{radius}).  We see that $R$ lies in
the range $8-82$~AU while $D$ lies in the range $33-165$~AU.  Binary
separation would be 0.''09 to 0.''47.  Such a separation may be
possible to resolve with the VLA or adaptive optics in the
near-infrared.

\subsection{Expected precession in binary systems with nonaligned jets}
\label{sec:obs2}

In the systems presented in this section, misaligned ``binary'' jets
have been observed. Since it is very unlikely that one single source
can drive two jets with very different orientations, it is assumed
that each of the outflows originates from its own component of a
binary system. In some cases a binary has actually been resolved, in
other cases observations only allow us to put an upper limit on the
separation of the hypothetical binary.  Since the outflows are not
parallel, it is probable that the disks which surround these sources
are themselves misaligned.  Therefore, at least one of these disks is
not in the orbital plane and should precess.  We evaluate here the
precession period $T$ and give an estimate of the lengthscale
$\lambda= v/T$ over which the outflows should 'wiggle' as a result of
this precessional motion.  Since in general we do not know from which
member of the binary each jet originates, we will here again consider
mass ratios $0.5 \le M_s/M_p \le 2$, unless otherwise specified.  We
will also take $M_p=0.5$~M$_{\sun}$ unless otherwise specified, and
$\cos \delta=1$ (the results can be scaled for different values of
$\delta$ since $\omega_p \propto \cos \delta$).

\noindent {\bf{T~Tau}} is a binary located in Taurus, at a distance of
140~pc.  Observations show that two almost perpendicular jets
originate from this system (B\"ohm~\& Solf~1994).  A disk of estimated
radius $R \sim 27$--67~AU has been resolved around the visible
component, T~Tau~N (Akeson, Koerner~\& Jensen~1998).  Here we assume
that the disk around T~Tau~N is not in the orbital plane.  We take
$D=102$~AU (Ghez~et al.~1991) and for $R$ the observational values
reported above.  We fix $M_p=0.7$~M$_{\sun}.$ Since we are interested
in the precession of the jet emanating from the primary, we consider
$0.5 \le M_s/M_p \le 1$.  We then get $T \sim 5,000$--$4 \times
10^4$~years.  Since $v=70$~km~s$^{-1}$ for the jet emanating from
T~Tau~N (Eisl\"offel \& Mundt 1998), $\lambda \sim 0.4$--3~pc for this
jet.  These values of $\lambda$ are larger than the scale over which
the jet has been observed so far (which about 0.1~pc), so that a
bending rather than 'wiggling' may be detectable in that case.

\noindent {\bf{HH1 VLA 1/2}} is located at a distance of 480~pc.  The
two sources VLA~1 and~2, separated by 1,400~AU, drive the two
misaligned jets HH~1--2 and HH~144, respectively (Reipurth~et
al.~1993).  We assume that this system is bound and noncoplanar, and
that tidal truncation has operated such that $2 \le D/R \le 4$ (note
however that, in such a young and wide system, $D/R$ may actually be
significantly larger).  We also assume that the angle between the line
of sight and the orbital plane is 45$\arcdeg$, so that $D \simeq
1,980$~AU.  Then $T \sim 4 \times 10^5$--$4 \times 10^6$~years.  Since
$v \sim 200$~km~s$^{-1}$ for both flows (Eisl\"offel, Mundt~\&
B\"ohm~1994), we get $\lambda \sim 77$--870~pc.  Since $T$ is probably
comparable to the age of the system, a bending rather than wiggling of
the jet may be expected on a scale of a few pc.  We note that
'wiggling' or bending of the jets has been suggested on the current
observed scale, which is about 0.5~pc for both jets (Reipurth~et
al.~1993).  This clearly cannot be due to interaction between VLA~1
and~2, but it may be the sign that this system contains more sources.
This is supported by the existence of at least two more outflows with
different orientations (Eisl\"offel~et al.~1994).

\noindent {\bf{HH~111 IRS}:} Perpendicular to the HH~111 jet, located
at a distance of 480~pc in Orion, is another outflow called HH~121
(Gredel~\& Reipurth~1993; Davis~et al.~1994).  We assume that the
source of these two outflows is a binary.  From the unresolved central
source in VLA observations (Rodr\'{\i}guez~1997) we infer an upper
limit on the separation $D$ of about 0.''1, or 48~AU.  By adopting
$D=48$~AU and $2 \le D/R \le 4$, we find that $T \sim
1,000$--8,000~years.  Since $v \sim 350$~km\,s$^{-1}$ for HH~111
(Reipurth, Raga~\& Heathcote~1992), $\lambda \sim 0.5$-- 6~pc for this
jet.  The lowest of these values is close to the extent over which the
jet has been observed so far, which is 0.45~pc in projection
(Reipurth~1989). We note that tidal effects in a putative binary
system have been invoked by Gredel~\& Reipurth~(1993) as a possible
cause of the asymmetry of HH~121.

\noindent {\bf{HH~24 SVS63}:} The region of HH~24, located in Orion at
a distance of 480~pc, contains several highly collimated outflows
(Eisl\"offel~\& Mundt~1997) and a hierarchical system of four or even
five young stars. The sources SSV~63E and SSV~63W are separated by
about 4,500~AU in projection onto the sky (Davis et al. 1997).
SSV~63W is itself a binary.  On the images taken by Davis et
al. (1997) and given to us, we have measured that its projected
separation is 920~AU.  We find that SSV~63E is probably a triple
system: the projected separation is 350~AU between SSV~63E~A and~B and
975~AU between SSV~63E~A and~C.  We will take these projected
separations as indicative values of the actual separations.  At least
two outflows with very different orientations originate from
SSV~63E. These are HH~24~G (Mundt, Ray~\& Raga~1991) and HH~24~C/E
(Solf~1987; Eisl\"offel \& Mundt 1997) , which extend over 0.2 and
about 1~pc, respectively. SSV~63W is the source of another
parsec--scale outflow, HH~24~J (Eisl\"offel \& Mundt 1997).  Here
again we fix $2 \le D/R \le 4$.  The velocities of the jets, in
km~s$^{-1}$, are about 140, 180 (if we assume the radial and
tangential velocities to be similar), 370 and 50 for HH~24~J, HH~24~G,
HH~24~C/E and for the redshifted lobe HH~24~E, respectively (Jones~et
al.~1987).  The different interactions within the two systems then
lead to $\lambda$ between 5 and 556~pc, which is at least one order of
magnitude larger than the extent over which the jets have been
observed so far.  This indicates that a bending rather than a
'wiggling' would be more likely to be observed.

\noindent {\bf{L1551~IRS5}} is located in Taurus, at a distance of
150~pc.  It has been suggested that two jets could be emanating from
this source (Moriarty--Schieven~\& Wannier~1991, Pound~\&
Bally~1991). Furthermore, Rodr\'{\i}guez~et al.~(1998) have shown that
this system is a binary with separation $45$~AU and they have resolved
two circumstellar discs for which they infer $R \sim 10$~AU.  Their
results are also consistent with the presence of two outflows which
appear to be misaligned (see their Fig.~2).  We take $D$ between 45
and 63~AU, corresponding to an angle between the orbital plane and the
line of sight in the range 0--45$\arcdeg$, and we fix $R=10$~AU.  Then
$T \sim 4,000$--$5 \times 10^4$~years.  Since for the jet which has
been unambiguously observed $v \sim 200$~km~s$^{-1}$ (Sarcander,
Neckel~\& Els\"asser~1985), we derive $\lambda \sim 1$--10~pc.  The
projected extent over which the jet has been observed so far is about
1~pc (Moriarty--Schieven~\& Snell~1988).  Therefore, even though a
full period may not yet be seen, 'wiggling' could already appear in
the observations.  We note that the outflow as observed appears to be
very complex, and it may well be seen to precess.

\section{Discussion and Summary}
\label{sec:disc}

In this {\it Letter} we have considered several protostellar systems
where either a precessing jet or at least two misaligned jets have
been observed. In the case where a jet is seen to precess (or rather
interpreted as precessing), we have assumed it originates from a disk
which is tidally perturbed by a companion on an inclined orbit, and we
have evaluated the parameters of the binary system. For Cep~E,
V1331~Cyg and RNO~15--FIR, we found the separation to range from a few
AU up to 160~AU and the disk size to be between 1 and 80~AU. These
numbers correspond to what is expected in pre--main sequence binaries
(see Mathieu~1994 and references therein).  We note that larger
separations for this range of disk sizes would be associated with
longer precession timescales, and thus it would be more difficult to
detect the precessional motion over the observed lengthscale of the
jet. A bending rather than 'wiggling' would be expected in that case.

In the case where misaligned jets have been observed, we have assumed
or used the fact that the source of these outflows is a binary, and we
have calculated the precession frequency that would be induced by
tidal interactions in the binary and the lengthscale over which the
jets should 'wiggle' as a result of this precessional motion.  For
T~Tau, HH1~VLA~1/2 (which may actually be a hierarchical system) and
HH~24~SVS63, it may be possible to detect a bending of the jets rather
than 'wiggling' on the current observed scale of the jets. In
HH~111~IRS and L1551~IRS5 (assuming there are indeed two misaligned
outflows in this system), 'wiggling' may be detected on the projected
scale (0.5--1~pc) over which the jets have been observed.

Our results are consistent with the existence of noncoplanar binary
systems in which tidal interactions induce jets to precess. Some of
the predictions of this {\it Letter} could be tested observationally
in a near future.

\acknowledgements 

We thank Chris Davis for supplying us promptly with the images of
SSV63.  We acknowledge the Isaac Newton Institute for hospitality and
support during its programme on the Dynamics of Astrophysical Discs,
when this work began.  CT is supported by the Center for Star
Formation Studies at NASA/Ames Research Center and the University of
California at Berkeley and Santa Cruz.

\newpage

%{\centerline{FIGURE CAPTION}}

%\figcaption[fig.ps]{The binary separation $D$ vs. the disk radius $R$
%in~AU, as calculated from equation~(\ref{radius}), for Cep~E ({\it
%a}), V1331~Gyg ({\it b}) and RNO~15--FIR ({\it c}). The different
%curves correspond to $M_s/M_p=0.5$ ({\it solid line}), 1 ({\it dotted
%line}) and 2 ({\it dashed line}).  $D/R$ is kept in the range 2--4 and
%the symbols indicate $D/R=4$ ({\it crosses}), 3 ({\it squares}) and 2
%({\it triangles}). \label{fig1}}

\end{document}